\documentclass[aip,jcp,amsmath,amssymb,reprint,longbibliography,groupedaddress]{revtex4-1}

\usepackage[utf8]{inputenc}
\usepackage{braket}
\usepackage{amsmath}
\usepackage{bbm}
\usepackage{appendix}
\usepackage{amsfonts}
\usepackage{comment}
\usepackage{cancel}
\usepackage{gensymb}
\usepackage{bbold}
\usepackage{graphicx}
\usepackage{MnSymbol}
\usepackage{mathtools}
\usepackage{tabularx}
\usepackage{array}
\usepackage{accents}
\usepackage{bm}
\usepackage{xcolor}
\usepackage{placeins}
\usepackage{enumitem}   
\usepackage{amsthm}
\usepackage{soul}
\usepackage[justification=Justified,singlelinecheck=off]{caption}
\usepackage{tikz}
\usepackage{quantikz}

\newcolumntype{Y}{>{\centering\arraybackslash}X}

\usepackage[colorlinks, linkcolor=blue]{hyperref}
\usepackage[nameinlink, capitalise]{cleveref}

\newcommand{\nn}{\nonumber} 
\newcommand\numeq[1]%
{\stackrel{\scriptscriptstyle(\mkern-1.5mu#1\mkern-1.5mu)}{=}}

\newcommand{\tcb}{\textcolor{blue}}

\newcommand{\bs}{\boldsymbol}

\newcommand{\diag}{{\mathrm{diag}}}

\newcommand\commin[1]{\iffalse #1 \fi}

\newcommand{\mr}{\mathrm}

\newcommand{\mc}{\mathcal}

\newcommand{\QL}{\ensuremath{\mathrm{QL}}}
\newcommand{\NQL}{\ensuremath{{N_\mathrm{QL}}}}
\newcommand{\Ng}{\ensuremath{{N_{\mathrm{G}}}}}

\newcommand{\Ntot}{\ensuremath{{N_{\mathrm{tot}}}}}

\newcommand{\e}{\ensuremath{\,\mathrm{e}}}

\newcommand{\da}{\ensuremath{\downarrow}}
\newcommand{\ua}{\ensuremath{\uparrow}}
\newcommand{\bda}{{\pmb{\downarrow}}}
\newcommand{\bua}{{\pmb{\uparrow}}}

\newcommand{\CNOT}{\mathrm{CNOT}}

\usepackage{chngcntr}

\newcommand\norm[1]{\left\lVert#1\right\rVert}

\numberwithin{equation}{section}

\usepackage{etoolbox}
\makeatletter
\renewrobustcmd{\cref}{\@osmcref{cref}}
\renewrobustcmd{\Cref}{\@osmcref{Cref}}
\def\@osmcref#1#2{%
	\begingroup
	\ifcsundef{r@#2}
	{}
	{\expandafter\expandafter\expandafter\expandafter\expandafter
		\expandafter\expandafter\def
		\expandafter\expandafter\expandafter\expandafter\expandafter
		\expandafter\expandafter\@osmcref@name
		\expandafter\expandafter\expandafter\expandafter\expandafter
		\expandafter\expandafter{%
			\expandafter\expandafter\expandafter
			\@thirdoffive\csname r@#2\endcsname}}%
	\ifcsundef{r@#2@cref}
	{}
	{\cref@gettype{#2}{\@osmcref@type}}%
	\ifboolexpr{not test {\ifdefvoid{\@osmcref@name}}
		and (test {\ifdefstring{\@osmcref@type}{thm}}
		or test {\ifdefstring{\@osmcref@type}{lemma}})}
	{\nameref{#2} (\@cref{#1}{#2})}
	{\@cref{#1}{#2}}%
	\endgroup
}
\makeatother

\crefname{axioms}{axiom}{axioms}
\Crefname{axiom}{Axiom}{Axioms}

\begin{document}
\title{Encoding quantum-like information in classical synchronizing dynamics}
\author{Graziano Amati, Gregory D. Scholes}
\affiliation{Department of Chemistry, Princeton University, Princeton, NJ 08544, USA.\looseness=-1}

\date{\today}

\begin{abstract}
In previous work, we introduced a formalism that maps classical networks of nonlinear oscillators onto a quantum-like Hilbert space. We demonstrated that specific network transformations correspond to quantum gates, underscoring the potential of classical many-body systems as platforms for quantum-inspired information processing.
In this paper, we extend this framework by systematically identifying the classical dynamics best suited for this purpose. Specifically, we address the question: Can the collective steady state of a classical network encode signatures of quantum information?
We prove that the answer is affirmative for a special class of synchronizing many-body systems—namely, a complex-field extension of the Kuramoto model of nonlinearly coupled classical oscillators. 
Through this approach, we investigate how quantum-like entangled states can emerge from classical synchronization dynamics.
\end{abstract}

\maketitle

\section{Introduction}\label{sec:intro}

Classical synchronization is an ubiquitous phenomenon occurring in the dynamics of a wide range of physical and biological systems. From schools of fish and flocks of birds to pedestrians walking in lockstep, collective feedback mechanisms can give rise to robust collective states in classical many-body systems. \cite{birnir2017,okeeffe2017}

Classical synchronizing networks are known to be  robustness against noise and disorder—whether arising from static imperfections in the network or from dynamic fluctuations. \cite{scholes2023,gates,gupta2014}
This inherent resilience naturally leads to the question: Could these networks be exploited as platforms for information storage and processing?
This question has been investigated across various frameworks in the literature.
For example, stochastic networks of nanomagnets have been explored as a resource for implementing error-resilient invertible Boolean logic. \cite{camsari2017}
Networks of synchronizing oscillators from spin-torque and insulator-to-metal-transition devices have been studied as platforms for classical non-Boolean logic. \cite{raychowdhury2019,shukla2014}
The collective behavior of oscillator networks has also been leverages to solve computational problems naturally encoded in their structures, such as graph coloring and the Max-Cut problem. \cite{crnkic2020,bashar2020}
This naturally raises the question of whether classical networks can encode information in a quantum-like (QL) way.
Quantum-inspired approaches have extensively demonstrated their ability to capture key aspects of quantum computation using classical resources. For instance, tensor-network–based algorithms have achieved remarkable accuracy in quantum simulations on classical hardware. \cite{tindall2024,huggins2019}
Similarly, quasiclassical methods for quantum dynamics have proven effective in accurately describing quantum decoherence and thermalization while maintaining a sub-quantum computational complexity. \cite{mannouch2023,CHIMIA,ellipsoid,GQME,thermalization}

Conducting a large-scale analysis of the relationship between classical synchronizing networks and quantum information requires addressing several key challenges: Which types of classical dynamical systems, when structured as networks, are best suited for encoding QL information? Can these networks be engineered to implement quantum-inspired logic? What are the fundamental scalability limits of this approach as system size increases?
In this work, we systematically investigate these questions, evaluating both the feasibility and scalability of this framework.

A broad understanding of synchronization dynamics in classical many-body systems has been developed through studies of the Kuramoto model—a network of classical nonlinear oscillators that synchronize and reach collective consensus across a broad range of initial conditions and system parameters. \cite{kuramoto1975,rodrigues2016,arenas2008} 
This model naturally connects with graph theory, particularly in identifying network topologies that optimize synchronization in oscillator dynamics. \cite{scholes2022,scholes2025dyn}
Erdős-Rényi random graphs and expander families (e.g., k-regular graphs) are particularly relevant in this context.
Spectral analysis of their adjacency matrices reveals that a small subset of eigenvalues stands out distinctly from a densely populated bulk spectrum. The correspondent eigenvectors encode information on the synchronization patterns of
a system of classical oscillators assigned to the graph's nodes. \cite{scholes2023,scholes2020,scholes2024,product,lubotzky2012,abdalla2024}

In this work, we extend our previous studies within this framework by utilizing a complex embedding of the standard Kuramoto model, first introduced in Ref.~\onlinecite{muller2021,muller2023}. As we will demonstrate, this model further strengthens the connection between classical synchronization and QL information processing.
We explore how the spectral properties of specific classical networks shape their long-time dynamics and demonstrate that their synchronizing steady states can encode distinct signatures of quantum information.

\begin{figure*}
\centering
\includegraphics[width=\linewidth]{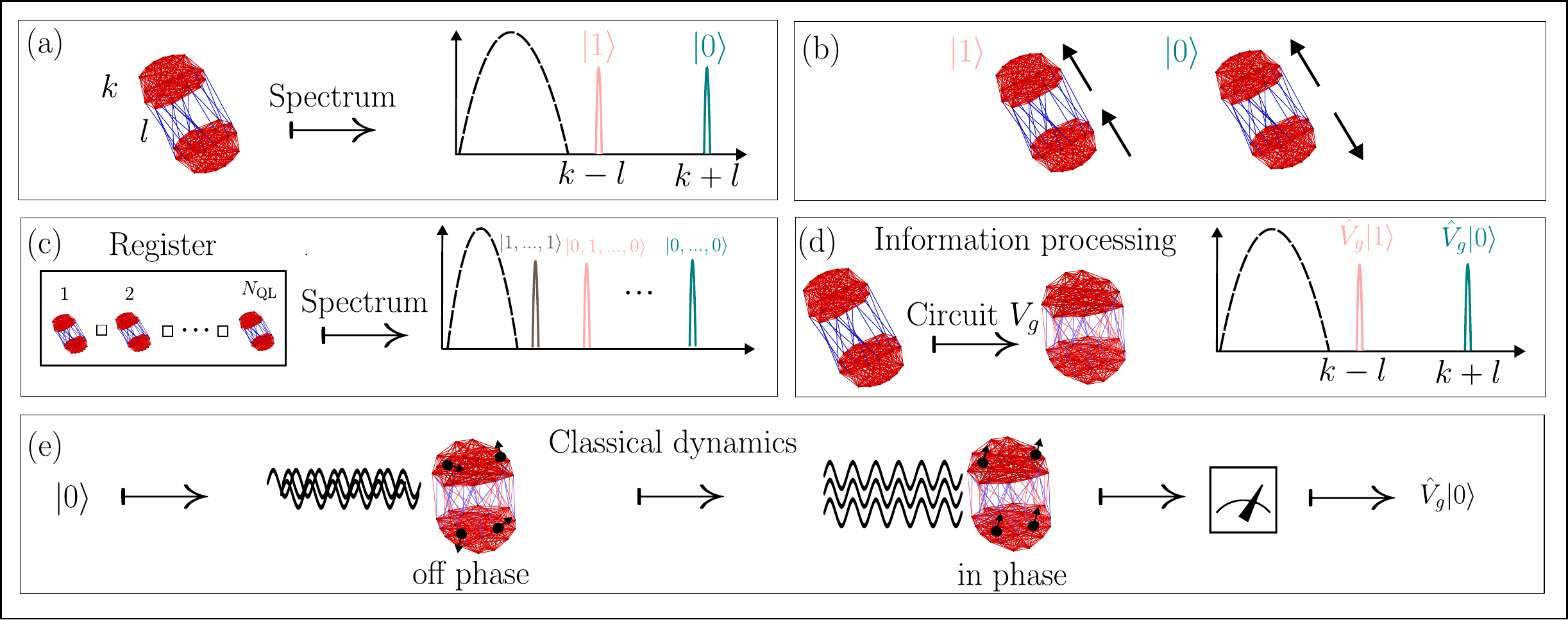}\caption{Illustration of the fundamentals of QL information processing as developed in recent work as in the present paper. Panel (a): Our approach considers networks of classical oscillators with two-body interactions governed by specially constructed regular graphs. Specifically, we consider two densely connected subgraphs, each with valency $k$ (red), mutually interacting through a cross-subgraph connectivity matrix with valency $l$ (blue). The adjacency matrix of these graphs exhibits a spectrum featuring two isolated eigenvalues at $k \pm l$. The corresponding eigenvectors are mapped onto the computational basis of a QL bit Hilbert space.
Panel (b): The two emergent states correspond to synchronized configurations, where the two subgraphs oscillate either in phase or in opposite phases relative to each other.
Panel (c): The Cartesian product ($\square$) of $\NQL$ resources gives rise to a set of emergent states that is isomorphic to the product basis of the Hilbert space of an arbitrary number of qubits.
Panel (d): By transforming the edges of the graphs, we can implement transformations corresponding to arbitrary quantum gates.
Panel (e): This approach enables the implementation of arbitrary QL circuits by allowing a network of classical oscillators to synchronize through their intrinsic nonlinear dynamics. The classical configurations of the system, when measured after classical synchronization, preserve quantum information initially encoded in the structure of the graph.}\label{fig:illustration}
\end{figure*}

To help the reader connect this work with our previous results, we illustrate the fundamental concepts of our QL approach in \cref{fig:illustration}. Technical details on the formalism are extensively discussed in Refs.~\onlinecite{scholes2023,scholes2024,product,gates,scholes2025dyn}.
Panel (a) depicts two densely connected regular graphs (highlighted in red) interacting through a sparse connectivity matrix (blue edges). The spectrum of the adjacency matrix of the network reveals two distinct peaks (in pink and green), clearly separated from a broad band of densely populated eigenvalues (dashed line). The locations of these peaks are determined by the valencies \( k \) and \( l \) of the elementary subgraphs, namely the number of neighbors of each node. 
The eigenvectors corresponding to these peaks are referred to as \emph{emergent states}. We establish a mapping from these states onto the computational basis of the Hilbert space of a QL bit. 
As illustrated in panel (b), these states correspond to two synchronized configurations of the network, namely in-phase and opposite-phase oscillations between the two subsystems.
Panel (c) illustrates how a register of \( \NQL \) QL-bits can be constructed by taking the Cartesian product (\(\square\)) of a set of elementary classical networks. The emergent states of the resulting composite system are isomorphic to the product basis of a system of qubits.  
Panel (d) depicts how unitary operations can be applied to the system to emulate the action of an arbitrary set of quantum gates. Since these operations are unitary, the spectrum of the resource remains unchanged, while the emergent states undergo transformations that mimic an arbitrary set of gates. 
Panel (e) highlights the core contribution of this work. Here, we design a network to emulate a specific quantum circuit and allow its classical dynamics to synchronize, meaning the oscillator phases converge to a global, collective pattern.
We then demonstrate that a classical measure of the system’s steady state encodes information about a target gate operation applied to an arbitrary QL state. This result reveals that quantum information can be inherently stored and processed within a classical synchronizing system.
We apply this framework to the classical realization of entanglement, demonstrating how QL circuits can be designed to generate analogs of Bell states. Furthermore, we explore the broader implications of our approach and discuss possible extensions to more complex QL circuits and algorithms, along with a potential strategy to mitigate the exponentially increasing computational cost associated with these large classical systems.

This paper is organized as follows: In \cref{sec:QLinf}, we introduce the fundamentals of the QL formalism, summarizing key results from our recent work. \Cref{sec:sync_gs} examines the classical dynamics of the QL ground state, constructed from a generalized Kuramoto model that is particularly relevant to our framework, as originally introduced in Ref.~\onlinecite{muller2021}. In \cref{sec:circ_sync}, we extend our analysis to QL circuits and explore entanglement within a two-QL-bit Hilbert space. Finally, in \cref{sec:protocols}, we present a protocol for the experimental implementation of our formalism.

\section{Quantum-like information processing}\label{sec:QLinf}

In this section, we summarize key results from Ref.~\onlinecite{gates,product,scholes2024,scholes2025dyn}, to provide the background for a self-contained understanding of the formal analysis presented in this paper. 
In \cref{subsec:hilbert} we introduce a mapping from graphs describing classical networks resources to the Hilbert space of a system of QL bits. The connection between graph transformations and QL gates is then discussed in \cref{subsec:operators}. For additional technical details, interested readers are encouraged to consult the referenced works.

\subsection{Hilbert space}\label{subsec:hilbert}

Here, we discuss how a computational basis for QL information processing can be distilled from the adjacency matrix a network of classical oscillators.

Let us consider a graph whose adjacency matrix exhibits the following block structure:
\begin{equation}\label{eq:R1}
\mc R_\da(k,l) = \begin{bmatrix}
A(k) &  -C(l) \\
-C^\top(l) &  B(k)
\end{bmatrix}.
\end{equation}
Each block on the right-hand side corresponds to a regular graph with \( \Ng \) nodes and valencies \( k \) and \( l \), as specified. The latter denote the (constant) number of edges drawn from each node of a regular subgraph.
The arrow ($\da$) on the left-hand side of \cref{eq:R1} labels the eigenvector of the matrix associated to the largest eigenvalue.
Specifically, a spectral analysis of \cref{eq:R1} reveals that, for several choices of $k$ and $l$, two of the $2\Ng$ eigenvectors correspond to the largest and most isolated eigenvalues, namely  
$\lambda_\da = k + l$ and $ \lambda_\ua=k - l$, with corresponding eigenvectors
\begin{equation}\label{eq:psi_pm}
\Psi_\da =  \tfrac 1 {\sqrt 2}\begin{pmatrix}
a_1\\ 
-a_2
\end{pmatrix} ,  \hspace{6mm} 
\Psi_\ua = \tfrac 1 {\sqrt 2}\begin{pmatrix}
a_1\\ 
a_2
\end{pmatrix} .
\end{equation}
Note that, since all blocks in \cref{eq:R1} are regular, the entries of the vectors $ a_i =(1, 1, \ldots, 1)^\top/\sqrt{2\Ng}$ are uniform.
However, the notation used here allows for a more general scenario in which impurities and disorder in the network may break the exact regularity condition.
A mapping from the emergent eigenvectors to the abstract Hilbert space of a single qubit is defined by 
$ \Psi_\sigma \mapsto \ket{\sigma} $,
where $ \sigma \in \mc{S} = \{\da, \ua\} $.
\cite{gates}
As a notation, throughout this work, we establish the correspondence \( \Psi_\da \equiv \Phi_1 \) and \( \Psi_\ua \equiv \Phi_2 \).  
The other eigenvectors \( \Phi_l \), with $2 \le l \le \Ng$, correspond to all other non-emergent eigenvalues.  
In this paper, we adopt both conventions for denoting the eigenvectors, depending on whether it is necessary to highlight the role emergent states in a spectral decomposition.

\begin{figure*}
\begin{minipage}{0.3\linewidth}
\centering
\includegraphics[width=1.15\linewidth]{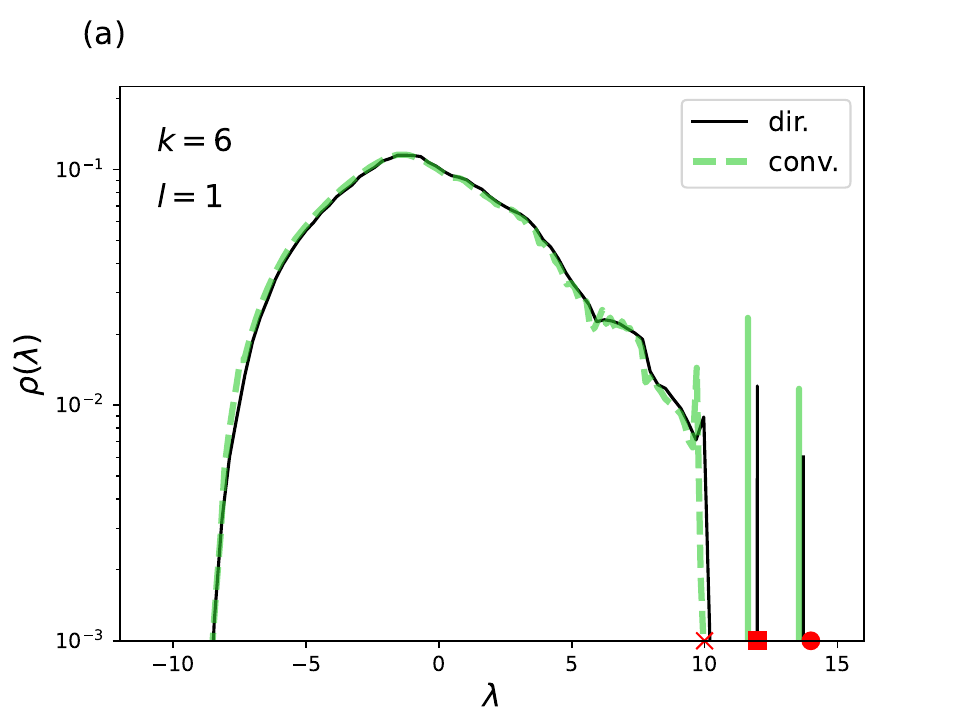}
\end{minipage}\hspace{0.33cm}
\begin{minipage}{0.3\linewidth}
\centering
\includegraphics[width=1.15\linewidth]{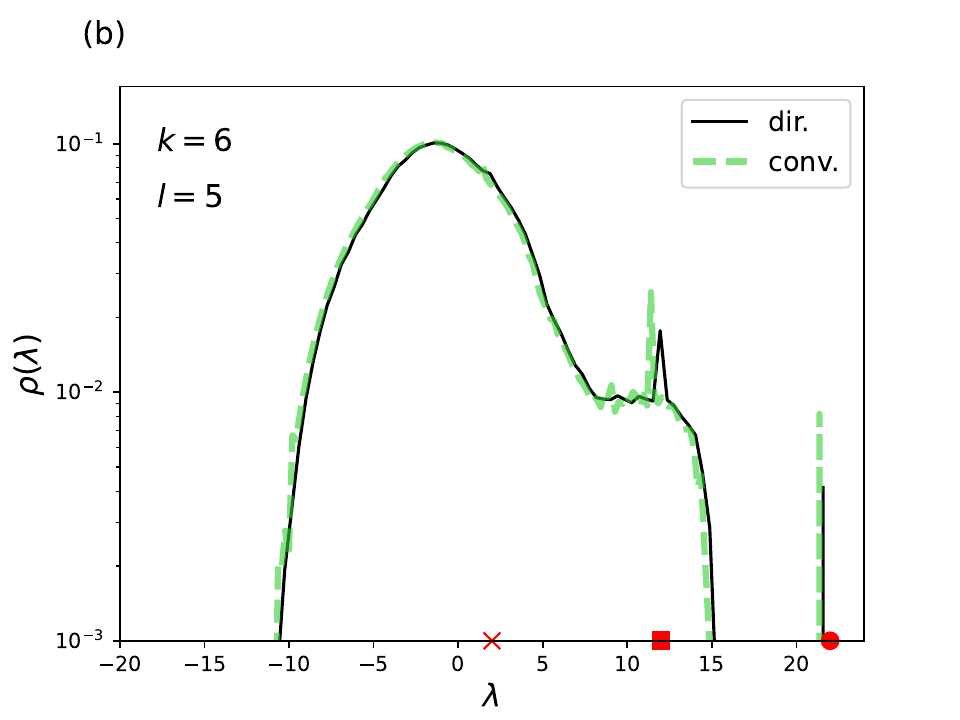}
\end{minipage}\hspace{0.33cm}
\begin{minipage}{0.3\linewidth}
\centering
\includegraphics[width=1.15\linewidth]{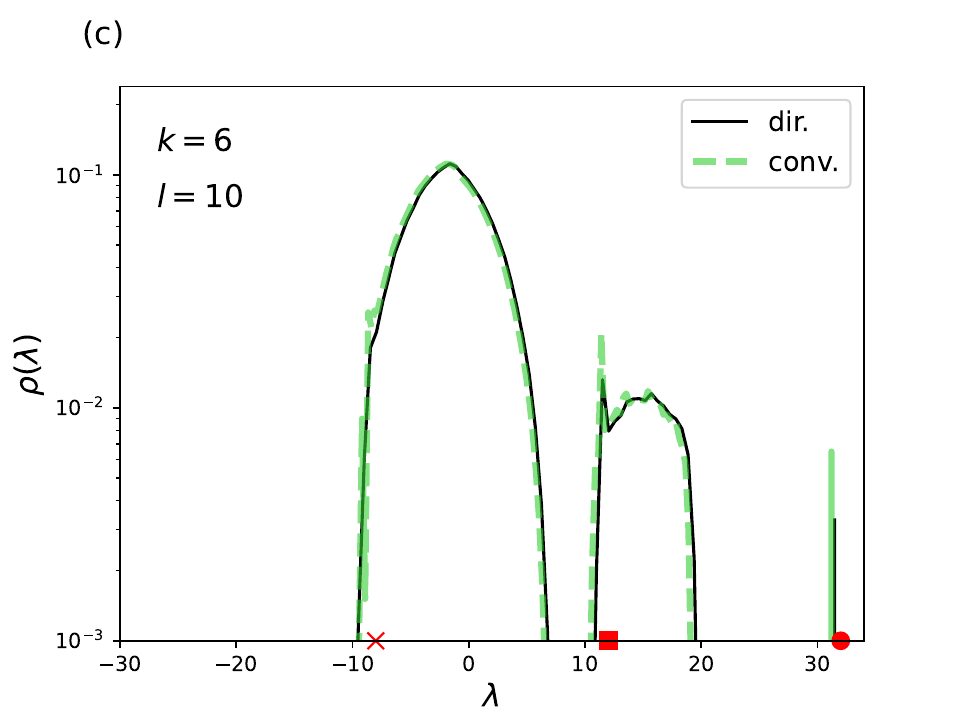}
\end{minipage}
\caption{Eigenvalue distribution for a system of two QL-bits, computed using both direct diagonalization (dir.) of \cref{eq:RNQL} and the convolution (conv.) of one-QL-bit spectra, as defined in \cref{eq:rho_conv}. The subgraphs forming the resources have identical valencies, $k_1 = k_2 = k$ and $l_1 = l_2 = l$. Red markers on the horizontal axis indicate the locations of the emergent eigenvalues at $2(k+l)$ (circle), $2k$ (square), and $2(k-l)$ (cross). Each single-QL-bit resource $\mc{R}^{(q)}$ consists of $\Ng = 12$ vertices, while $k$ and $l$ vary as specified in each panel. Each distribution is obtained by averaging over $N_{\mr{samp}} = 500$ random regular graphs.}
\label{fig:spectra}
\end{figure*}

In Ref.~\onlinecite{product}, we discussed how the structure of \cref{eq:R1} can be generalized to accommodate the correlations of a system of $\NQL$ QL bits.
There, we proposed to consider a network defined by the Cartesian product 
\begin{equation}\label{eq:RNQL}
\mc R_{\bda}(\bs k,\bs l)= \sum_{q=1}^\NQL \left[\mathbb{1}_{2\Ng}^{\otimes (q-1)}\right]\otimes \mc R_{\da}^{(q)}(k_q,l_q)\otimes  \left[\mathbb{1}^{\otimes (\NQL-q)}_{2\Ng}\right],
\end{equation}
where each $\mc{R}_\da^{(q)}$ is defined as in \cref{eq:R1}. The vectors $\bs{k}, \bs{l} \in \mathbb{N}^{\NQL}$, with entries $k_q$ and $l_q$ respectively, represent the set of valencies of all the elementary subgraphs.
The bold arrow on the left-hand side of \cref{eq:RNQL} is defined by $\bda = \{\da, \ldots, \da\} \in \bs {\mc S} = \{\da,\ua\}^\NQL$. 
The emergent eigenstates of \cref{eq:RNQL} are mapped onto the $2^\NQL$-dimensional product basis of a system of QL-bits, according to
\begin{equation}\label{eq:psi_sigma} 
\Psi_{\bs\sigma} = \bigotimes_{q=1}^{\NQL}  \Psi_{\sigma_q}^{(q)}\mapsto \bigotimes_{q=1}^{\NQL}  \ket {\sigma_q^{(q)}}\equiv \ket{\bm \sigma},
\end{equation}
where each single-QL-bit term $\Psi_{\sigma_q}^{(q)}$ is defined as in \cref{eq:psi_pm} and $\bs \sigma= \{ \sigma_1, \cdots, \sigma_{N_\QL} \}\in \bs{\mc S}$.
A spectral analysis of the Cartesian product in \cref{eq:RNQL} reveals that its eigenvalues are simply the sum of the contributions from individual QL bits.
In particular, the eigenvalues follow a distribution determined by their convolution
\begin{equation}\label{eq:rho_conv}
\rho(x) = \int_{-\infty}^{+\infty} \prod_{q=1}^{\NQL-1} \mr d x_{q}\; \rho^{(q)}(x_q) \rho^{(\NQL)}\left(x-\sum_{p=1}^{\NQL-1}x_p\right),
\end{equation}
where $\rho^{(q)}(x_q)$ denote the distributions of the spectra of the elementary resources in \cref{eq:R1}.
Numerical simulations support the validity of \cref{eq:rho_conv}, as illustrated in the three panels of \cref{fig:spectra}. There, we display the eigenvalue distribution of \( \mc{R}_{\bda} \) for \( \NQL = 2 \) and for different values of the elementary subgraph valencies, with \( k_1 = k_2 \equiv k \) and \( l_1 = l_2 \equiv l \).
We compare results from the direct (dir.) calculation of the spectrum of \cref{eq:RNQL}, and from the convolution (conv.) of two single-QL-bit spectral densities, as described in \cref{eq:rho_conv}.
To the best of the authors' knowledge, an exact analytical expression for the elementary distributions $\rho^{(q)}(x_q)$ in \cref{eq:rho_conv} has not yet been derived. Although the non-emergent part of the spectrum for sparse adjacency matrices can be approximated using the Wigner semicircle distribution, its convolution leads to an elliptic integral that lacks an exact solution. \cite{khalkhali2022,dumitriu2012,mckay1981,miller2008}
Given these mathematical challenges, a numerical analysis of the full spectra of this class of graphs remains a feasible and easily applicable approach. 
The subset of emergent eigenvalues can be instead be determined analytically, and given by
$ \lambda_{\bs{\sigma}} = \sum_{q=1}^{\NQL} \lambda_{\sigma_q} $.
These eigenvalues are integers of the form $i(k+l)+j(k-l)$, where the non-negative integers $i,j$ obey the constraint $i+j=\NQL$. We define $\lambda_{\bda}=\NQL(k+l)$ as the highest emergent eigenvalue $\lambda_{(\ua,\da,\da,\cdots, \da)}\equiv (\NQL-1)(k+l) + k-l$ as the second-highest term, and so on, down to the lowest emergent contribution $\lambda_{\bua}=\NQL(k-l)$. \cite{product}
The location of these special eigenvalues is marked in \cref{fig:spectra} by red markers, as specified in the caption.

Having established the foundations for mapping the state of classical networks onto a computational Hilbert space, we now turn to \cref{subsec:operators}. There, we summarize key results on constructing an isomorphism between classical network transformations and quantum gates, demonstrating how topological and dynamical properties of classical networks can be linked to quantum information processes.

\subsection{Gate operators}\label{subsec:operators}

In Ref.~\onlinecite{gates}, we proposed a mapping between standard quantum gates and unitary transformations applied to the adjacency matrix of a classical oscillator network arranged in graph structures.
For this purpose, we introduced a map $U_{\mr{cb}}$ transforming the emergent eigenvectors onto the standard computational basis (cb).
Given that all elementary subrgaphs in our resources [\cref{eq:R1,eq:RNQL}] are regular, this map can be expressed analytically by
\begin{equation}\label{eq:Ucb}
U_{\mr{cb}} =  [V_{\mr H}\otimes \mathbb 1_\Ng]^{\otimes \NQL},
\end{equation}
where
$
V_{\mr H} = \frac 1 {\sqrt 2}\begin{pmatrix}
1 & 1 \\
1 & -1
\end{pmatrix}
$ denotes the matrix representation of the Hadamard (H) gate in the computational basis.
For example, $U_{\mr{cb}}$ transforms the emergent eigenvectors of a single QL bit as 
\begin{equation}\label{eq:Ucb_1}
U_{\mr{cb}}\Psi_\da = \begin{pmatrix}  1_\Ng \\  0_\Ng \end{pmatrix}, \hspace{5mm} U_{\mr{cb}}\Psi_\ua = \begin{pmatrix}  0_\Ng \\  1_\Ng \end{pmatrix},
\end{equation}
where
\begin{equation}\label{eq:xM}
	x_M = (x,\cdots, x)/\sqrt{L} \in \mathbb C^{L}, \hspace{7mm}x \in \mathbb C,\; L \in \mathbb N.
\end{equation}
Let us consider now a quantum circuit described by $M$ gate operations $\bs g = \{g_1,\ldots, g_M\}$.
Be $V_{\bs g} = V_{g_M}\cdots  V_{g_1}$ the representation of this gate in the computational basis.
In Ref.~\onlinecite{gates} we defined by 
\begin{equation}\label{eq:Ug}
U_{\bs g} =  U_{\mr{cb}}^{-1} (V_{\bs g} \otimes  \mathbb 1_{\Ng}) U_{\mr{cb}} = U_{ g_M}\ldots U_{ g_1},
\end{equation}
the corresponding unitary map acting on QL states, which are built from linear combinations of the basis states in \cref{eq:psi_sigma}.
Note that the factorization in \( V_{\bs g} \) is preserved under the QL map, meaning that \( U_{\bs g} \) decomposes as \( U_{\bs g} = U_{g_M} \ldots U_{g_1} \).
To implement our QL approach in practice, we establish a direct connection between quantum gates and network transformations.
This framework defines rules for modifying the dynamics of classical synchronizing systems in a QL-consistent manner. A well-defined unitary map for this purpose is given by \cite{gates}
\begin{equation}\label{eq:R_to_Rg}
 \mc R_{\bda}\mapsto \mc R_{\bs g} = U_{\bs g} \mc R_{\bda}  U_{\bs g}^\dagger.
\end{equation}
We will show later in this paper that \cref{eq:R_to_Rg} transforms all eigenstates of a network resource consistently to the gate $U_{ \bs g}$ as defined in \cref{eq:Ug}.
Note that, here and in the following, we omit the explicit parametric dependence of $\mc R_\bda$ and $\mc R_{\bs g}$ on the valency vectors $\bs k$ and $\bs l$ for ease of notation. 

Building on our QL framework, we explore key questions in the following sections: How do the steady states of classical synchronizing dynamics relate to emergent states in graph structures? How can the mapping in \cref{eq:psi_sigma} from emergent states to Hilbert space be practically implemented? And what role do the QL gate operations in \cref{eq:R_to_Rg} play within this framework?
To address these questions, we first examine a synchronizing  system that is particularly well-suited for our QL approach. Specifically, we consider a complex embedding of the well-known Kuramoto model, a network of nonlinear oscillators evolving according to non-symplectic classical dynamics. \cite{kuramoto1975,scholes2020,muller2021,acebron2005}

\section{Ground-state synchronization}\label{sec:sync_gs}

In this section, we examine a classical synchronizing model designed to encode the ground state of a quantum information system within its collective steady-state dynamics.
In \cref{subsec:gs_dyn}, we present a closed-form solution for the time evolution of this model, as originally derived in Ref.~\onlinecite{muller2021}. 
In \cref{subsec:gs_ss}, we analyze how the long-time behavior of the system relates to the ground state of a qubit system. We extend this framework to QL information processing in the later \cref{sec:circ_sync}.
 
\subsection{Dynamics}\label{subsec:gs_dyn}

\begin{figure*}
\begin{minipage}{0.49\linewidth}
\centering
\includegraphics[width=0.9\linewidth]{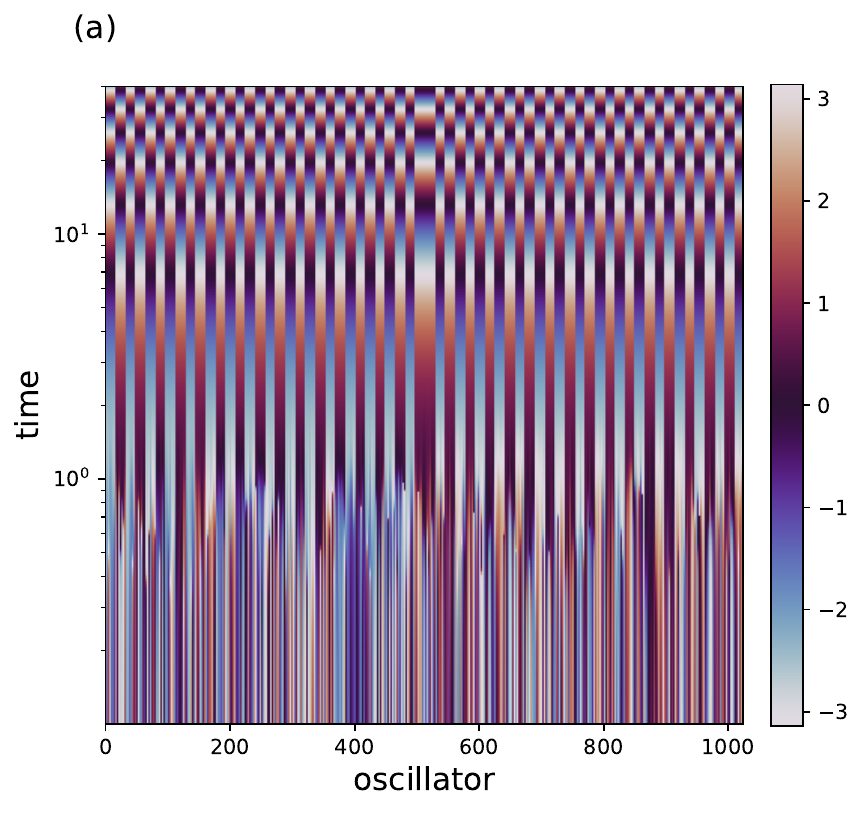}
\end{minipage}
\begin{minipage}{0.49\linewidth}
\centering
\includegraphics[width=0.9\linewidth]{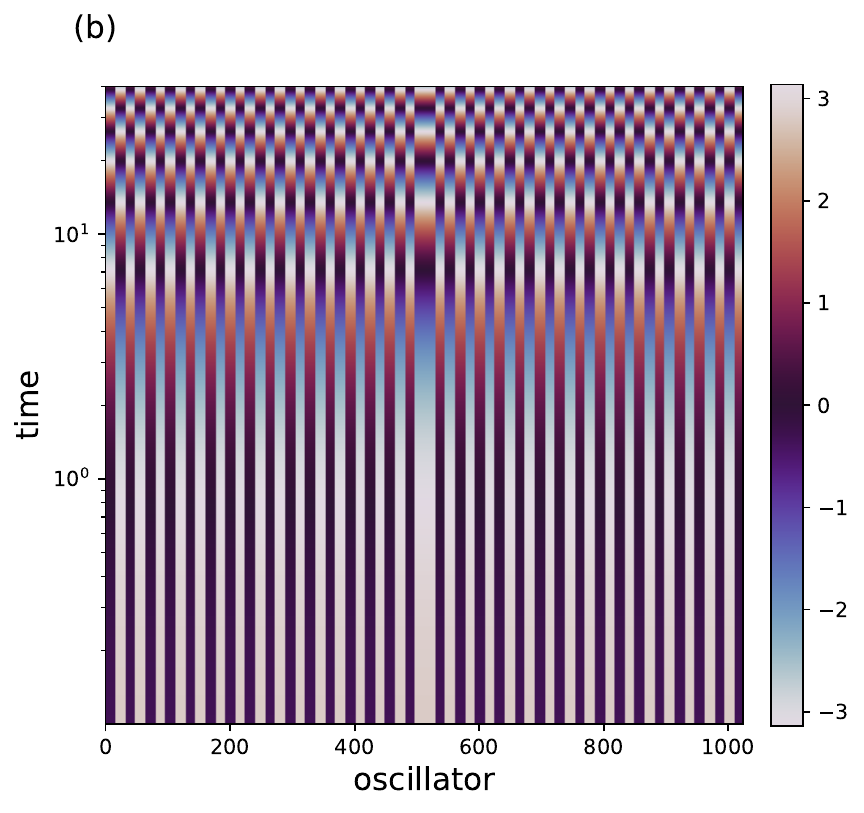}
\end{minipage}
\caption{Panel (a): Synchronization dynamics of the angles \( \theta(t) \) in the complex embedding of the Kuramoto model, as defined in \cref{eq:kuramoto}.  
The two-body interactions between oscillators are determined by a network structured according to the ground-state resource in \cref{eq:RNQL}.  
We consider a system of \( \NQL=2 \) QL bits, each constructed from subgraphs of \( \Ng=16 \) classical oscillators.  
The valencies of all diagonal and off-diagonal elementary subgraphs are fixed at \( k_1 = k_2 = 8 \) and \( l_1 = l_2 = 4 \), respectively.  
The initial angles of the oscillators are uniformly sampled from \( [0,2\pi) \).  
Finally, we set \( \overline{\omega} = 0.5 \) and \( K = 10 \). Panel (b): Emergent-state approximation of the steady-state dynamics of the system shown in panel (a).  }
\label{fig:theta_gs}
\end{figure*}

The Kuramoto model is a non-symplectic dynamical system consisting of nonlinearly coupled oscillators. 
The model is known for giving rise to classical synchronization across a broad range of initial conditions and system parameters. \cite{kuramoto1975,muller2021,kawamura2010}
Following the approach proposed in Refs.~\onlinecite{muller2021,muller2023}, we consider here a complex embedding of this model, defined by the equations of motion
\begin{align}
\dot \theta_l 
&= \omega_l -\frac{i K}{\Ntot}\sum_{m=1}^{\Ntot} \left[\mc R_{\bda}\right]_{lm}\e^{i(\theta_m-\theta_l)},\label{eq:kuramoto}
\end{align}
where $\Ntot=(2\Ng)^\NQL$. 
Here, a set of classical oscillators with frequencies $\theta_l$ interact nonlinearly through a complex coupling term, characterized by the global coupling strength $K$ and the adjacency matrix $\mathcal{R}_{\bda}$.
Note that the real part of \cref{eq:kuramoto} corresponds to the standard Kuramoto model.
As we will show later in this paper, the essential synchronization properties of the standard model are retained within this complex embedding.
A major advantage of \cref{eq:kuramoto} is that the complex-phase coupling allows for a closed-form solution of the time propagator.
This can be seen by expressing \cref{eq:kuramoto} in matrix form, as
\begin{align}
&\frac{\mr d}{\mr d t} \e^{i \theta} = \left[\diag (i\omega)+\frac{K}{\Ntot}\mc R_{\bda}\right] \e^{i\theta} \label{eq:kur_start}
\end{align}
or, more compactly as, $\dot{ x} = \mc D_{\bda} x$, where $ x = \e^{i\theta} = (\e^{i\theta_1},\ldots, \e^{i\theta_\Ntot})^\top$, while 
\begin{equation}\label{eq:D_da}
\mc D_{\bda} = \mr{diag}(i\omega)+\frac{K}{\Ntot}\mc R_{\bda}= i\overline \omega \mathbb 1_{\Ntot} + \frac{K}{\Ntot}\mc R_{\bda}
\end{equation}
denotes the generator.
To simplify our analysis, we fix here and in the following all oscillator frequencies to a unique common value, $\omega_j=\overline\omega >0 \;\;\forall \; j$.
In this case, the eigenvalue problems of $\mc R_{\bda}$ and $\mc D_{\bda}$ are identical, apart from a constant shift and rescaling of the eigenvalues.
Given that all terms in \cref{eq:RNQL,eq:D_da} commute, the propagator takes a simple factorized form, given by
\begin{align}\label{eq:expD_gs_R}
\e^{\mc D_{\bda} t } 
&=\e^{i\overline \omega t} \prod_{q=1}^\NQL\exp\Bigg\{\frac{K t}{\Ntot}  \mathbb{1}_{2\Ng}^{\otimes (q-1)}\otimes \mc R_{\da}^{(q)} \otimes  \mathbb{1}^{\otimes (\NQL-q)}_{2\Ng} \Bigg\}\nn\\
&=\e^{i\overline \omega t}\bigotimes_{q=1}^\NQL \e^{\frac{K t}{\Ntot} \mc R_{\da}^{(q)}}.
\end{align}
The spectral decomposition of the ground-state resource of each QL bit $(q)$ can be expressed as
\begin{align}\label{eq:Rspec}
\mc R_{\da}^{(q)} &= 
\sum_{l_q=1}^{2\Ng} \lambda_{l_q}^{(q)} \Phi_{l_q}^{(q)} \left[\Phi_{l_q}^{(q)}\right]^\dagger,
\end{align}
which leads to
\begin{align}
\e^{\mc D_\bda t}
&= \e^{i\overline\omega t}\bigotimes_{q=1}^\NQL \sum_{l_{q}=1}^{2\Ng}\e^{\frac{K t}{\Ntot}\lambda_{l_{q}}^{(q)}} \Phi_{l_q}^{(q)} \left[\Phi_{l_q}^{(q)}\right]^\dagger. \label{eq:expD_gs}
\end{align}
By convention, the sum in \cref{eq:Rspec} is defined in decreasing order for the eigenvalues of $\mc R_{\da}^{(q)}$.
These are all real, given that this adjacency matrix is itself real and symmetric.
Also, here and throughout the manuscript, we require that all eigenvectors define an orthonormal basis. 
According to the spectral theorem, this condition is satisfied as long as an orthonormalization procedure is applied exclusively within the subspaces that are found to be degenerate.
We define the initial state \( x_0^{(q)} \) of each subsystem by uniformly sampling the initial oscillator angle vector \( \theta_0^{(q)} \in [0,2\pi) \) and expressing it as \( x_0^{(q)} = \e^{i\theta_0^{(q)}} \).
The total initial state is defined by the Cartesian product,
\begin{align}\label{eq:x0_prod}
x_0 &= \bigotimes_{q=1}^\NQL x_0^{(q)} = \bigotimes_{q=1}^\NQL\sum_{{l_q}=1}^{2\Ng} c_{l_q}^{(q)} \Phi_{l_q}^{(q)}.
\end{align}
where
$c_{l_q}^{(q)} = \langle \Phi_{l_q}^{(q)}, x_0^{(q)} \rangle$.
With \cref{eq:expD_gs,eq:x0_prod}, we can express the solution of the Kuramoto model as
\begin{align}
x_{\bda}(t) &= \e^{\mc D_\bda t} x_0 
= \e^{i\overline\omega t}\bigotimes_{q=1}^\NQL \sum_{l_{q}=1}^{2\Ng}c_{l_q}^{(q)}\e^{\frac{K t}{\Ntot}\lambda_{l_{q}}^{(q)}} \Phi_{l_q}^{(q)} .  \label{eq:xt_gs}
\end{align}
The tensor product in \Cref{eq:xt_gs} indicates that, in the ground state, each single-QL-bit network defined by $\mc R^{(q)}_\da$ evolves independently of the other subsystems. 
The numerical solution of the angles $\theta_\bda(t)\equiv\angle x_\bda(t)$ is illustrated in panel (a) of \cref{fig:theta_gs}.
Specifically, we display the evolution of the angle vector \( \theta_\bda(t) = \{\theta_{\bda,1}(t), \ldots, \theta_{\bda,\Ntot}\} \) by plotting, over time (vertical axis), the angle of each oscillator, with individual oscillators labeled on the horizontal axis.
The model parameters are detailed in the figure caption.
At time \( t = 0 \), the oscillators are initialized with random angles, uniformly sampled from \( [0,2\pi) \). 
After a brief transient, synchronization emerges as the oscillators self-organize into \( 2\Ng \) subgroups, each consisting of \( 2\Ng \) oscillators, with adjacency groups oscillating in opposite phases.
This phase offset results from the negative sign in the off-diagonal coupling blocks of the corresponding correlation matrix, as defined in \cref{eq:R1}.  
The group structure aligns with the multiplicative effect induced by the tensor product in \cref{eq:expD_gs}.
Let us note that, unlike most studies on Kuramoto synchronization—where oscillators adjust their individual frequencies $\omega_l$ to achieve global coherence—the synchronization observed here can be considered ``weak.'' 
This is because all degrees of freedom already share the same intrinsic frequency, \( \omega_l = \overline{\omega} \).  
In our case, synchronization arises purely from the evolution of the oscillators' phases  
$\phi_l(t) = \theta_l(t) - \overline{\omega} t$,
which gradually converges to a single constant value over time.

The present analysis shows that synchronization naturally emerges within this complex embedding of the Kuramoto model.  
The long-time synchronization patterns are shaped by the network topology, which defines the two-body interactions between the classical oscillators.  
In the next \cref{subsec:gs_ss}, we examine the steady-state solution of these dynamics and explore its connection to the ground state of a QL-bit system.

\subsection{Steady state}\label{subsec:gs_ss}

An important goal at the foundation of this work is to establish a connection between long-time classical synchronization and quantum information.  
We begin addressing this problem by focusing on the QL analog of the ground state \(\ket{\bda}\).  
In our approach, this state is mapped onto \(\Psi_{\bda}\), which, by construction, is the leading eigenvector of \(\mathcal{R}_{\bda}\), with largest eigenvalue \(\lambda_{\bda}\). 
Here, we show that, at long times, the configurations of the oscillators arrange in a state that is approximately proportional to \(\Psi_{\bda}\).  
This result provides us with a protocol for encoding robust quantum information—emerging from synchronization—onto the steady-state solution of a classical many-body system.

From \cref{eq:xt_gs} we observe that the configuration vector $x_{\bda}(t)$ depends exponentially on the eigenvectors of the graph adjacency matrix.
The leading emergent eigenvalue, $\lambda_1^{(q)}\equiv \lambda_\da^{(q)}$, can be significantly larger than all other components for a suitable choice of the system parameters and of time.
In this case, we can approximate the full expansion of $x_{\bda}(t)$ in \cref{eq:xt_gs} with only the ground-state contribution, and define the approximate configuration vector
\begin{equation}\label{eq:bar_xt_gs}
\tilde x_{\bda}(t) = \e^{i\overline \omega t} \bigotimes_{q=1}^\NQL c_1^{(q)} \e^{ \frac{ \lambda_\da^{(q)}K t}{\Ntot}}\Psi_\da^{(q)}=\e^{i\overline \omega t} \prod_{q=1}^\NQL  c_1^{(q)} \e^{ \frac{ \lambda_\da^{(q)}K t}\Ntot }\Psi_{\bda},
\end{equation}
where, as established in \cref{subsec:hilbert}, $\Phi_1^{(q)}\equiv \Psi_\da^{(q)}$ and $ \Psi_\bda = \bigotimes_{q=1}^\NQL \Psi_\bda^{(q)}$.
The solution of \cref{eq:bar_xt_gs} is shown in panel (b) of \cref{fig:theta_gs}.
We observe a good agreement between the long-time steady state of \cref{eq:xt_gs} and \cref{eq:bar_xt_gs}.
This result demonstrates that, in this system, the QL ground state can be directly encoded in the renormalized limit of the long-time dynamics, namely:
\begin{equation}\label{eq:x_to_Psi}
\boxed{x_\bda(t)  \approx \tilde x_\bda(t)\propto \Psi_\bda, \hspace{7mm} t \gg 1.}
\end{equation}
We examine later in this paper how to optimize the model parameters in order to ensure the validity of the approximation in \cref{eq:x_to_Psi}.

In the next \cref{sec:circ_sync}, we discuss how the results from this section can be generalized when QL gate maps \cref{eq:R_to_Rg} are applied to the matrix \( \mc R_{\bda} \).  
This enables us to develop a strategy for leveraging classical synchronization to encode quantum information.

\section{Circuit synchronization}\label{sec:circ_sync}

A key result from \cref{sec:sync_gs} is \cref{eq:x_to_Psi}, which establishes a direct connection between classical synchronizing dynamics and the ground state of a qubit system.
In this section, we explore whether this correspondence can be extended to arbitrary gate transformations and circuits. We find that this is indeed the case [see \cref{eq:bar_xgt_state}].
Our analysis demonstrates that properly designed synchronizing classical dynamics do not dissipate quantum information. Instead, they preserve it and effectively encode it across their degrees of freedom.

\subsection{Dynamics}\label{subsec:circ_dyn}

Let us consider a quantum circuit described by a sequence of $M$ gate operations \( \bs g = \{g_1, \dots, g_M\} \). 
The correspondent of \cref{eq:D_da} in this case is defined by
\begin{equation}\label{eq:Dg}
\mc D_{\bs g} = U_{\bs g} \mc D_{\bda}  U_{\bs g}^\dagger = i\overline{\omega} \mathbb{1}_{\Ntot} + \frac{K}{\Ntot} \mc R_{\bs g}.
\end{equation}  
Single-QL-bit gates, defined by maps of the form
\begin{equation}
U_g = \mathbb{1}_{2\Ng}^{\otimes(q-1)} \otimes U_{g}^{(q)} \otimes \mathbb{1}_{2\Ng}^{\otimes(\NQL - q)},
\end{equation}
preserve the structure of the Cartesian product in \cref{eq:RNQL}. 
In this case, the dynamics of each oscillator group maintain their independent, factorized structure as in \cref{eq:expD_gs}, with the only difference that the graph of a single QL-bit (q) is transformed to
\begin{equation}
\mathcal{R}^{(q)} \mapsto U_{g}^{(q)} \, \mathcal{R}^{(q)} \, \left[ U_{g}^{(q)} \right]^\dagger.
\end{equation}
However, this factorization no longer holds once two-QL-bit gates are applied.  
As shown later in the context of entanglement, these transformations replace certain identity operators in \cref{eq:RNQL} with nontrivial maps, breaking the Cartesian-product structure of the ground state network.  
Since any higher-dimensional gate can be decomposed into a sequence of one- and two-QL-bit operations, our analysis can be effectively restricted to these two fundamental cases. \cite{barenco1995}  
As a consequence of this symmetry breaking, the solution of the network dynamics can no longer be expressed using \cref{eq:bar_xt_gs}, requiring a new formulation.  
We derive this by starting from the spectral decomposition of the circuit-transformed resource
\begin{equation}\label{eq:Rg}
\mc R_{\bs g} = 
\sum_{l=1}^\Ntot \lambda_{l} \left[ U_{\bs g}\Phi_{ l}\right] \left[ U_{\bs g}\Phi_{ l}\right]^\dagger,
\end{equation}
where the scalar indices 
\begin{equation}\label{eq:l_row_maj}
l = \sum_{q=1}^\NQL (2\Ng)^{q-1}l_q
\end{equation}
are the one dimensional ordering of the vector indices $\bs l = \{l_1,\ldots, l_{2\Ng}\}$.
Similarly, we defined
$\lambda_{l} = \sum_{q=1}^\NQL \lambda_{l_q}^{(q)}$, where $\{\lambda_{l_q}^{(q)}\}_{l_q=1}^{2\Ng}$ are the eigenvalues of $\mc R_\da^{(q)}$.
\Cref{eq:Rg} is based on the observation that the spectrum of a matrix is invariant under unitary transformations, while the eigenvectors change accordingly.
Similarly to the ground-state result in \cref{eq:expD_gs_R}, we can now express the system propagator as
\begin{align}
\e^{\mc D_{\bs g} t} 
&=\e^{i\overline \omega t} \exp\left\{\frac{K t}{\Ntot} \sum_{l =1}^\Ntot\lambda_{l} \left[ U_{\bs g}\Phi_{ l}\right] \left[ U_{\bs g}\Phi_{l}\right]^\dagger\right\}\nn\\
&=\e^{i\overline \omega t}\sum_{l=1}^\Ntot \e^{ \frac{ \lambda_{l}K t}{\Ntot}}\left[ U_{\bs g}\Phi_{l}\right] \left[ U_{\bs g}\Phi_{ l}\right]^\dagger,\label{eq:expD_g}
\end{align}
where we noticed that, for any unitary map $U_{\bs g}$,
\begin{align}
\langle U_{\bs g} \Phi_l, U_{\bs g} \Phi_{m}\rangle = \langle \Phi_{l}, \Phi_{m} \rangle 
= \delta_{ l,m}.
\end{align}
Therefore, the propagation of the initial state $x_0$ [from \cref{eq:x0_prod}] under the dynamics generated by \cref{eq:expD_g} is given by
\begin{align}
x_{\bs g}(t) &=\e^{\mc D_{\bs g} t}x_0 
= \e^{i\overline \omega t}\sum_{l,m=1}^\Ntot c_m \e^{ \frac{ \lambda_{ l}K t}\Ntot} U_{\bs g}\Phi_{l}
\langle U_{\bs g}\Phi_{l}, \Phi_{m}'\rangle,\label{eq:xgt}
\end{align}
where $m$ is the scalar representation of $\bs m =\{m_1,\ldots, m_{2\Ng}\}$ [as in \cref{eq:l_row_maj}],  $c_m = \prod_{q=1}^{\NQL} c^{(q)}_{m_q}$ and $\Phi_m '= \bigotimes_{q=1}^\NQL \Phi_{m_q}^{(q)}$.
The exponential dependence of the eigenvalues in \cref{eq:xgt} suggests, similarly to \cref{eq:bar_xt_gs}, that it might be possible to infer information on the leading emergent state of this system, $U_{\bs g}\Phi_1\equiv U_{\bs g}\Psi_{\bda}$, from the steady-state dynamics of specially prepared networks.
We discuss this aspect in detail in the next \cref{subsec:entanglement}, where we apply the present formalism to analyze a circuit generating an entangled state.

\subsection{Application to entanglement}\label{subsec:entanglement}

\begin{figure*}
\begin{minipage}{0.49\linewidth}
\centering
\includegraphics[width=0.9\linewidth]{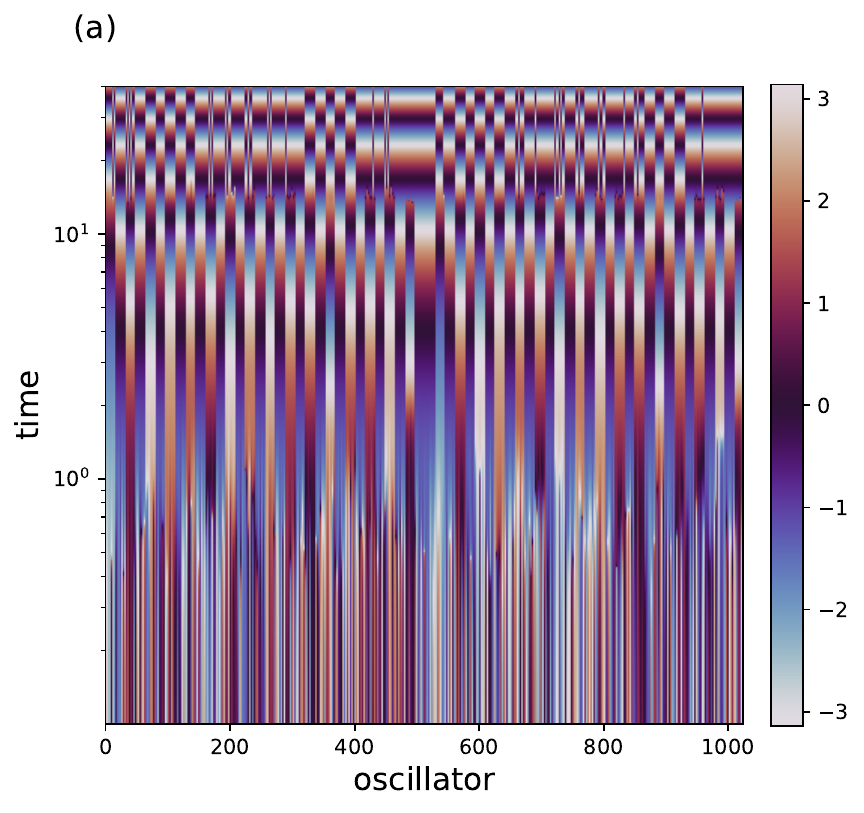}
\end{minipage}
\begin{minipage}{0.49\linewidth}
\centering
\includegraphics[width=0.9\linewidth]{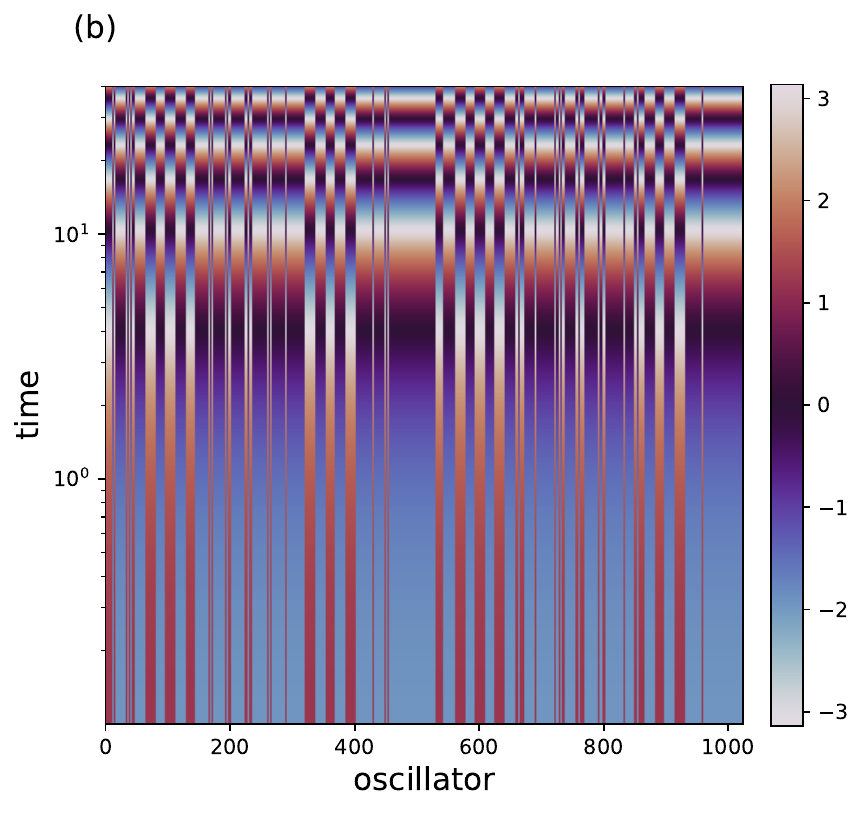}
\end{minipage}
\caption{Panel (a): Time evolution of the angular vector in the Kuramoto model, obtained from \cref{eq:xgt}, with the connectivity matrix \( \mathcal{R}_{\bm{g}} \) defined by the Bell gate in \cref{eq:R_Bell}.  
All system parameters remain identical to those in \cref{fig:theta_gs}, except for the structure of the adjacency matrix.  
Panel (b): Time evolution of the emergent state dynamics for the same system as in panel (a), computed from \cref{eq:bar_xgt}.}
\label{fig:theta_Bell}
\end{figure*}

Here, we apply the formalism from  \cref{subsec:circ_dyn} to study a two-QL-bit system of classical oscillators encoding the circuit $\bm{g} = \bm{g}_{\mathrm{B}} = \{\mathrm{H}, \CNOT\}$.
This map transforms the ground state to the QL correspondent of the Bell state $\ket{\Phi_+} = \tfrac{1}{\sqrt{2}} (\ket{00} + \ket{11})$.
The QL map defining the circuit is
\begin{align}\label{eq:U_Bell}
U_{\bs g_{\mr B}}  
&= \left[\mc P_{\da}^{(1)} \otimes \mathbb 1_{2\Ng} +  \mc P_{\ua}^{(1)}  \otimes U^{(2)}_{\mr X}\right]\left[ U_{\mr H}^{(1)}\otimes \mathbb 1_{2\Ng}\right],
\end{align}
where we introduced the projection operators
\begin{equation}\label{eq:P_pm_gates}
\mc P_\da^{(q)} = \frac 12 \left[\mathbb 1_{2\Ng} + U_{\mr Z}^{(q)}\right], \hspace{5mm} \mc P_\ua^{(q)} = \frac 12 \left[\mathbb 1_{2\Ng} - U_{\mr Z}^{(q)}\right],
\end{equation}
which are mutually orthogonal and complementary:
\begin{align}
\mc P_\sigma^{(p)} \mc P_{\sigma'}^{(p)}=\delta_{\sigma\sigma'}\mc P_{\sigma'}^{(p)},\hspace{8mm}
\mc P_\da^{(p)} + \mc P_\ua^{(p)}=\mathbb 1_{2\Ng}.
\end{align}
By expanding the general definition of QL gates \cref{eq:Ug} with $V_{\mr Z} = \begin{pmatrix} 0 & 1 \\ 1 & 0 \end{pmatrix}$ and with \cref{eq:Ucb}, we obtain
\begin{equation}\label{eq:P_pm_mat}
\mc P_\ua^{(q)} = \frac 12 \begin{pmatrix}
\mathbb 1_{2\Ng} & \mathbb 1_{2\Ng} \\
\mathbb 1_{2\Ng} & \mathbb 1_{2\Ng} 
\end{pmatrix}
, \hspace{5mm} 
\mc P_\da^{(q)}  =\frac 12   
\begin{pmatrix}
\mathbb 1_{2\Ng} & -\mathbb 1_{2\Ng} \\
-\mathbb 1_{2\Ng} & \mathbb 1_{2\Ng} 
\end{pmatrix}.
\end{equation}
In this circuit, the initial resource \cref{eq:RNQL} for $\NQL=2$ is transformed to 
\begin{align}
\mc R_{\bs g_{\mr B}} &= U_{\bs g_{\mr B}} \mc R  U_{\bs g_{\mr B}}^\dagger   
= \sum_{\sigma \in \mc S} \left(\mc P_\sigma^{(1)}\mc R_{\mr H}^{(1)}\mc P_{\sigma}^{(1)}\right)\otimes \mathbb 1_{2\Ng} \nn\\
&\quad + \sum_{\substack{\sigma, \sigma'\in \mc S\\\sigma\neq \sigma'}} \left(\mc P_\sigma^{(1)}\mc R_{\mr H}^{(1)}\mc P_{\sigma'}^{(1)}\right)\otimes U_{\mr X}^{(2)} \nn\\
&\quad +\mc P_\da^{(1)}\otimes \mc R^{(2)}+
 \mc P_\ua^{(1)}\otimes \mc R_{\mr X}^{(2)}.\label{eq:R_Bell}
\end{align}
The correspondent of \cref{fig:theta_gs} for the Bell circuit is shown in \cref{fig:theta_Bell}. 
The results are displayed in a similar format, with the horizontal axis representing the \( (2\Ng)^2 \) oscillators and the vertical axis corresponding to time.
Compared to the ground-state dynamics, the steady-state of the full dynamics [in panel (a)] observed here obeys a less symmetric synchronization pattern, reflecting the inhomogeneous structure of the two-body network from \cref{eq:R_Bell}. 
The observed dynamics can be classified into three distinct regimes. Initially, a noisy phase emerges due to the random initial conditions of the phase oscillators. This is followed by a transient quasi-synchronized state, which gradually evolves into a long-term steady state characterized by a distinct, non-symmetric synchronization pattern.
Following the same route as in \cref{sec:sync_gs}, we investigate now to what extent the steady-state dynamics can be accurately described by a single emergent-state term. 
From \cref{eq:xgt}, this contribution is defined by
\begin{equation}\label{eq:bar_xgt}
\tilde x_{\bs g}(t) =  \e^{\left(i\overline \omega  + \frac{\lambda_{\bda}K }{\Ntot}\right)t}U_{\bs g}\Phi_{\bda}\sum_{m=1}^\Ntot c_m 
\langle U_{\bs g}\Psi_\bda, \Phi_{m}'\rangle,
\end{equation}
where we identified $\lambda_1=\lambda_{\bda}$ as the largest emergent eigenvalue, with corresponding eigenvector $U_{\bs g} \Psi_\bda$.
A preliminary evaluation of the accuracy of the approximation \( x_{\bs g}(t) \approx \tilde{x}_{\bs g}(t) \), can be assessed by the direct comparison between \cref{eq:xgt,eq:bar_xgt}, show in panel (a) and (b) of \cref{fig:theta_Bell}, respectively. 
As previously observed for the ground-state dynamics, retaining only the leading emergent state provides a satisfactory description of the final synchronized state.
The result implies that the identity
\begin{equation}\label{eq:bar_xgt_state}
 \boxed{x_{\bs g}(t) \approx \tilde x_{\bs g}(t)  \propto U_{\bs g}\Psi_{\bda}, \hspace{7mm} t \gg 1}
\end{equation}
holds for this model system.
\Cref{eq:bar_xgt_state} is a central result of our paper, as it establishes a direct connection between classical synchronization dynamics and quantum information.
This finding demonstrates that classical synchronization can be utilized for quantum information storage and processing.
In particular, if \cref{eq:bar_xgt_state} holds, quantum gate operations can be pre-encoded in the initial network structure.
Gate transformations of QL states can then be extracted through a classical measurement of the oscillators' state once they reach a steady synchronized state.

\begin{figure}
\centering
\includegraphics[width=1\linewidth] {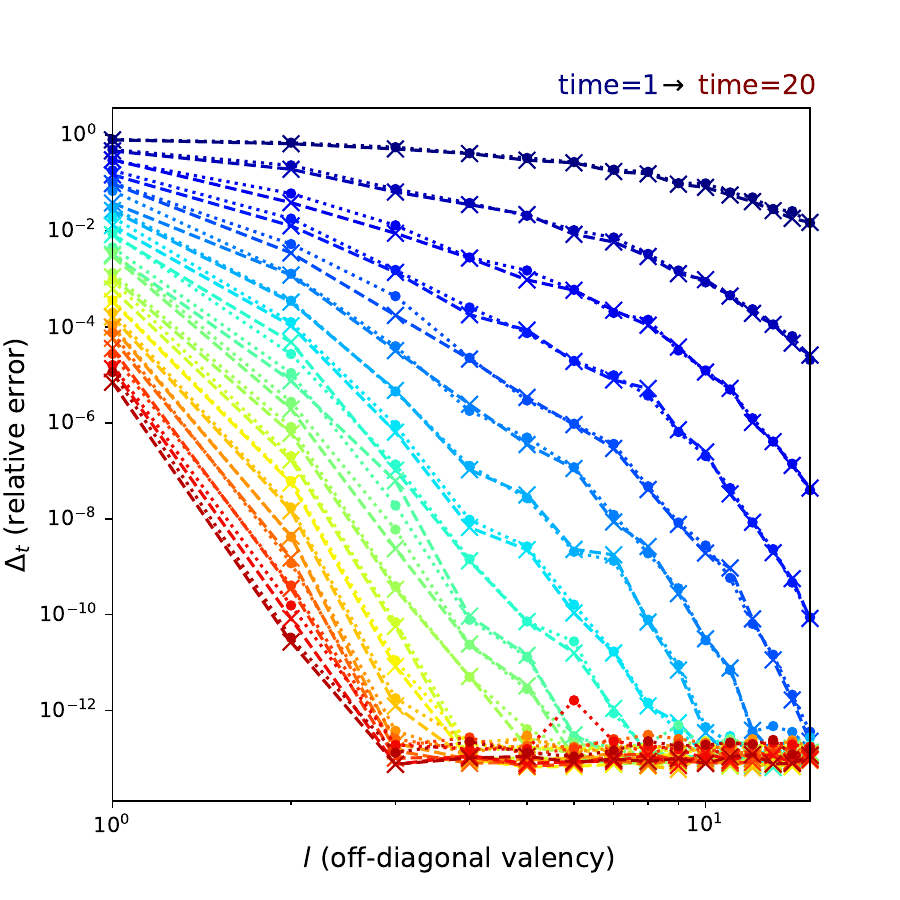}
\caption{Relative error between the exact solution of the oscillator dynamics and its emergent-state approximation, as defined in \cref{eq:Delta_t}.  
We consider a system of two QL bits, each constructed from subgraphs with diagonal valencies \( k_1 = k_2 = 4 \) and varying off-diagonal valencies \( l_1 = l_2 = l \), as shown on the horizontal axis.  
Each color represents a different time \( t \), ranging from \( t = 1 \) (dark blue) to \( t = 20 \) (dark red).  
The remaining system parameters are set to \( \overline{\omega} = 0.5 \) and \( K = 10 \).  
Results are averaged over \( N_{\mr{samp}} = 100 \) initial conditions for \( x_0 \), as described in \cref{subsec:gs_dyn}.}
\label{fig:error}
\end{figure}

We conclude this section by examining the conditions under which the approximation in \cref{eq:bar_xgt_state} remains well-defined.  
As shown in \cref{fig:spectra}, in a two-QL-bit system, the leading emergent eigenvalue, \( \lambda_{\bda} = 2(k+l) \), becomes increasingly separated from the rest as \( l \) grows. 
This suggests that increasing the valency \( l \) can improve the accuracy of the emergent-state approximation.  
This hypothesis is further supported by \cref{fig:error}, where we analyze the relative error
\begin{equation}\label{eq:Delta_t}
\Delta_t = \frac{\norm{x_{\bs g}(t) - \tilde x_{\bs g}(t)}}{\norm{x_{\bs g}(t)}}
\end{equation}
as a function of the valency \( l_1 = l_2 = l \) of the off-diagonal graphs used to construct both QL bits, and for different time values, as indicated by the color code.
In the figure, dotted lines with dots represent results from the ground-state circuit, while dashed lines with crosses correspond to the Bell circuit.  
As expected, the figure shows that as both the valency \( l \) and time increase, the accuracy of the approximation improves monotonically at an exponential rate.
Additionally, we observe that the relative errors in both cases are nearly identical, with only minimal discrepancies.  
This aligns with the fact that our approximation depends solely on the structure of the resource spectrum. This structure remains unchanged by any specific unitary transformation applied to the graph adjacency matrix—whether it be the identity operator, the Bell circuit, or any other QL circuit.  

In conclusion, our analysis demonstrates that QL information can be extracted from the long-time dynamics of classical resources, provided that the approximation in \cref{eq:bar_xgt_state} is sufficiently accurate.
This condition can be ensured by appropriately tuning the off-diagonal valencies of the elementary network resources. 
In the final \cref{sec:protocols}, we explore the implications of this finding and present a general framework for implementing our approach experimentally.

\section{Quantum-like information protocols}\label{sec:protocols}

We conclude this paper by presenting a general protocol for implementing our approach to QL information processing in both computational and experimental settings.
While the exact method described here faces scalability challenges due to the high cost of constructing large networks with many QL bits, we outline a strategy in \cref{app:comput_adv} that can be explored in future work to mitigate this ``curse of dimensionality.''
\\

QL theories can be applied by following these steps:

\begin{enumerate}[label=\arabic*, ref=\arabic*]
\item\label{item:1} Sample a set of \( \NQL \) elementary random regular graphs, \( A_i(k_q) \) (\( i = 1,2 \)) and \( C(l_q) \), from an initial estimate of the valencies \( \bs l = \{l_1, \dots, l_\NQL\} \) and \( \bs k = \{k_1, \dots, k_\NQL\} \). These graphs are then used to build single-QL-bit resources, \( \mc R_{\da}^{(q)}(k_q, l_q) \), with structure defined as in \cref{eq:R1}.
\item\label{item:2} Compute their Cartesian product to generate the ground-state resource as described in \cref{eq:RNQL}.  
\item\label{item:3} Repeat \ref{item:1} and \ref{item:2} to optimize the valency vectors \( \bs k \) and \( \bs l \) of the adjacency matrices, ensuring that the leading emergent state is well isolated from the rest of the spectrum. This guarantees that the emergent-state approximation \( {x}_{\bs g}(t) \approx \tilde{x}_{\bs g}(t) \) is well defined for the chosen network [see \cref{fig:spectra,eq:xgt,eq:bar_xgt}]. 
\item Apply a unitary transformation to the initial resource corresponding to a given quantum circuit, as defined in \cref{eq:R_to_Rg}. 
In an experimental setting, this requires constructing a Kuramoto model with tunable couplings and phases between oscillators.  
Various potential experimental implementations are explored in Ref.~\onlinecite{gates}.
\item  Let the classical network evolve until it reaches a collective synchronized state.  
Classical molecular dynamics simulations can help optimize system parameters to ensure synchronization,  
specifically the single-oscillator frequency \( \overline{\omega} \), the global coupling constant \( K \),  
and the sampling distribution of the initial frequencies \( \theta_l(0) \).
\item A readout of the long-time configuration vector of the oscillators accurately represents the state \( U_{\bs g} \Psi_{\bda} \), corresponding to the desired gate transformation applied to the ground state.  
\end{enumerate}

This protocol builds on the observation that a carefully designed classical system can preserve quantum information in its long-time dynamics, as long as quantum logic is  encoded in the two-body correlations between its degrees of freedom.  
By leveraging synchronization, classical many-body systems can be designed to efficiently store and process quantum information.

\section{Conclusions}

In this paper, we built upon our recent work linking classical synchronization dynamics to quantum information.  
Our approach extends the standard Kuramoto model of nonlinearly synchronizing oscillators through a complex-embedded formulation, originally proposed in Ref.~\onlinecite{muller2021}.  
We adopted this model for its ability to encode quantum information within the steady states of its synchronized dynamics.  
We demonstrated that arbitrary quantum circuits can be mapped onto the two-body correlations of model. Classical dynamics preserve this quantum information, which ultimately manifests in the synchronized steady states of the network.  
Two key results of this work, \cref{eq:bar_xt_gs,eq:bar_xgt}, explicitly define and establish this connection.
As an initial validation of our approach, we applied our framework to a network encoding the quantum ground state and to a two-QL-bit circuit generating an entangled state.  
However, our methodology is formulated in general terms and is fundamentally applicable to arbitrary quantum gates and circuits.

The number of edges in the Cartesian product graph from \cref{eq:RNQL} [and related gate transformations in \cref{eq:R_to_Rg}] grows exponentially with the number of QL bits, imposing a natural limit on the feasible system size for computational and experimental studies.
Nevertheless, even relatively small networks—within current computational and experimental capabilities—are sufficient to validate key proof-of-concept QL algorithms, as demonstrated in this work. In particular, optimized high-performance classical simulations can extend QL algorithm testing to systems with up to $\NQL\lesssim 10$ QL bits. \cite{stegailov2015}
A promising strategy for improving scalability exploits the fact that steady-state synchronization is governed by lower-rank approximations of the full network, as outlined in \cref{app:comput_adv}. This suggests that smaller, systematically designed graphs can preserve the most relevant synchronization properties of larger networks. Further enhancements could incorporate Laplacian-based spectral clustering for dimensionality reduction while retaining essential spectral and topological features. \cite{ding2024,wang2025} We aim to explore these approaches in future work to minimize computational complexity.
Despite challenges in scalability, the potential of processing quantum information within a classical framework remains significant. Classical resources are generally easier to implement and manipulate compared to genuine quantum systems.

We expect that our mapping from quantum to classical synchronization extends beyond quantum information processing and can also be applied to the study of quantum dynamics. More specifically, a key question we aim to explore in future work is the structure of classical synchronizing dynamics in a system where unitary gates [as defined in \cref{eq:Ug}] are replaced by the QL correspondent of quantum time evolution operators. 
This extension naturally connects to quasiclassical approaches to quantum dynamics, which have demonstrated high accuracy in capturing inherently quantum-mechanical phenomena—such as detailed balance and decoherence—by modeling quantum dynamics through classical equations of motion. \cite{GQME, thermalization, ellipsoid, CHIMIA, mannouch2023}
In particular, encoding an open quantum subsystem into robust emergent states of classical synchronizing networks could offer a promising strategy for mitigating and delaying the onset of decoherence and thermalization in open quantum systems. \cite{breuer2002,qubit}

Qudits have emerged as a compelling alternative to qubit-based quantum computing. 
Their advantages include reduced circuit depth, minimizing decoherence and gate errors, as well as improved resource efficiency and fault tolerance through better error management. \cite{wang2020,pavlidis2021,deller2023}
By leveraging the increased complexity of underlying basis graphs, our approach can naturally extend to qudit systems, and we are actively exploring these possibilities. This generalization aligns with our complementary goal of extending the formalism to simulate quantum dynamics. In particular, mapping graph correlations onto qudits instead of qubits offers a promising framework for studying quantum systems with arbitrary number of quantum levels.

Our theory is designed to be compatible with a variety of experimental setups. 
In particular, realizations of the Kuramoto model have been demonstrated across multiple experimental platforms, including LC circuits, \cite{allam2006,barbi2021,maffezzoni2015}, spin-torque oscillators, \cite{raychowdhury2019,belanovsky2013,kanao2019} and metal-to-insulator transition devices. \cite{kim2009,parihar2015,raychowdhury2019} 
These implementations support the assessment of the experimental feasibility of our approach and enable the construction of network resources suitable for testing small-to-medium-scale QL algorithms.

Large-scale open questions in our work concerns the intrinsic limitations of QL theories. 
Even with access to infinitely large classical networks, can this formalism fully recover quantum information? 
What are the precise connections between our QL notion of entanglement, nonlocality, and Bell inequalities? 
Can these phenomena be understood within the framework of hidden variable theories?
We aim to investigate these foundational questions in future work. A promising direction is to establish a broad connection between our approach and the mathematical and philosophical foundations of QL theories, particularly those developed in the context of generalized probabilistic models. \cite{khrennikov2010,khrennikov2019,khrennikov2022}

\section{Acknowledgments}

The authors would like to thank Dr. Debadrita Saha for useful discussions. This research was funded by the National Science Foundation under Grant No. 2211326 and the Gordon and Betty Moore Foundation through Grant GBMF7114.

\begin{appendix}

\section{Computational benefits of the emergent-state approximation}\label{app:comput_adv}

In this appendix, we analyze the computational advantages of the emergent-state approximation [from \cref{eq:bar_xt_gs,eq:bar_xgt}] in terms of resource efficiency and performance. This discussion underscores its potential as a practical strategy for reducing storage and processing demands in large systems of QL bits, both in computational simulations and experimental implementations.

The spectral decomposition of a network representing a circuit $\bs g$ is given by  
\begin{equation}
\mc R_{\bs g} = W_{\bs g} \Lambda W_{\bs g}^{-1} \in M_{\Ntot\times \Ntot}(\mathbb{C}),
\end{equation}
where $W_{\bs g}$ is the matrix that diagonalizes $\mc R_{\bs g}$, and $\Lambda$ is a diagonal matrix containing its eigenvalues.  
Note that, as discussed in \cref{subsec:circ_dyn}, the eigenvalues matrix $\Lambda$ is independent of the specific unitary gate transformation $\bs g$ applied to the resource.  
The emergent-state approximation studied throughout the paper relies on storing only the leading element of $\Lambda$.  
This corresponds to the upper-left corner of the matrix, as we order in this paper the eigenvalues in decreasing order.
In the following, we examine the more general case where the first $\tilde{N}_l \leq \Ntot$ diagonal elements of $\Lambda$ are retained, and we analyze the computational cost associated with storing and processing this reduced adjacency matrix.  
In this case, we define by
\begin{equation}\label{eq:tildeR_g}
\tilde{\mc R}_{\bs g} \approx \tilde W_{\bs g} \tilde \Lambda \tilde W_{\bs g}^{-1}
\end{equation}
the reduced eigenvalue problem, where $\tilde \Lambda \in M_{\tilde N\times \tilde N}(\mathbb C)$  involves the first $\tilde N$ nonzero elements on the diagonal, and similarly $\tilde W_{\bs g}\in M_{\Ntot,\tilde N}(\mathbb C)$ is a matrix storing in its columns the first $\tilde N$ eigenvectors of $\mc R_{\bs g}$.
Given that only the first  $\tilde N$ eigenvalues and eigenvectors of $\mc R_{\bs g}$ are needed in \cref{eq:tildeR_g}, iterative methods such as the Lanczos algorithm (for Hermitian matrices) \cite{cullum2002} or the Arnoldi method (for general matrices) \cite{frommer2013} can be used to solve this reduced eigenvalue problem. 
Their diagonalization cost depends on whether the given matrix is sparse or dense, as summarized in \cref{table:cost}.
\begin{table}[h]
    \centering
        \renewcommand{\arraystretch}{1.5} 

    \begin{tabular}{|c|c|c|}
        \hline
        & \textbf{Diagonalization} & \textbf{Storage } \\
        \hline
        \textbf{Full decomposition} & $\mathcal{O}(N_{\text{tot}}^3)$ & $\mathcal{O}(N_{\text{tot}}^2)$ \\
        \hline
        \textbf{Partial (sparse)} & $\mathcal{O}(N_{\text{tot}} \tilde{N})$ & $\mathcal{O}(N_{\text{tot}} \tilde{N})$ \\
        \hline
        \textbf{Partial (dense)} & $\mathcal{O}(N_{\text{tot}}^2 \tilde{N})$ & $\mathcal{O}(N_{\text{tot}} \tilde{N})$ \\
        \hline
    \end{tabular}
    \caption{Comparison of computational costs for different decomposition methods of QL network resources. Here $\tilde N \le \Ntot$ denotes the fraction of eigenvales and eigenvectors calculated over a total of $\Ntot=(2\Ng)^\NQL$ states. The ``Diagonalization'' column refers to the computational complexity of solving the eigenvalue problem, while the ``Storage'' column represents the memory requirements for storing the reduced representation.}
    \label{table:cost}
\end{table}
Thus, if \(\tilde N \ll \Ntot\), using iterative diagonalization approaches can significantly reduce the computational cost compared to full decomposition of the network resources.

\end{appendix}

\FloatBarrier
%

\end{document}